\documentclass{article}
\usepackage{spconf,amsmath,graphicx,hyperref}

\usepackage[inkscapearea=page]{svg}

\usepackage{mathtools}
\usepackage{hyperref}
\usepackage{amssymb}
\usepackage{bm}
\usepackage{graphicx}
\usepackage{float}
\usepackage{amsmath}
\usepackage{tikz}
\usepackage{bbm}
\usepackage{orcidlink}
\usepackage{multirow}
\usetikzlibrary{positioning, backgrounds, fit, shapes.arrows}
\usetikzlibrary{decorations.markings}
\usepackage{amsthm}
\usepackage[table]{xcolor}
\usepackage{cuted}
\usepackage{algorithmicx}
\usepackage{algorithm}
\usepackage{algpseudocode}
\usepackage{gensymb}
\usepackage[caption=false,font=footnotesize]{subfig}
\usepackage{enumitem}
\usepackage{caption}
\usepackage{subcaption}
\usepackage{amssymb}
\usepackage{pifont}

\tikzset{block/.style = {draw, fill=white, rectangle,
		minimum height=3em, minimum width=2cm},
	input/.style = {coordinate},
	output/.style = {coordinate},
	pinstyle/.style = {pin edge={to-,t,black}}
	radiation/.style={{decorate,decoration={expanding waves,angle=90,segment   length=4pt}}}
	
}
\usepackage{smartdiagram}

\tikzstyle{block} = [draw, rectangle, minimum height=2em, minimum width=2em]
\tikzstyle{sum} = [draw, circle,minimum width=0.1 cm]
\tikzstyle{input} = [coordinate]
\tikzstyle{output} = [coordinate]
\tikzstyle{dummy} = [coordinate]
\tikzstyle{pinstyle} = [pin edge={to-,thin,black}]
\usetikzlibrary{positioning, fit, arrows.meta}
\usetikzlibrary{positioning}
\usetikzlibrary{shapes,arrows}
\tikzstyle{frame_cyan} = [thick, draw=blue, solid,inner sep=0.3em]
\tikzstyle{frame_red} = [thick, draw=red, solid,inner sep=0.3em]
\tikzstyle{frame_green} = [thick, draw=green, solid,inner sep=0.3em]

\usepackage{tikz}
\usepackage{xcolor}
\definecolor{fc}{HTML}{1E90FF}
\definecolor{h}{HTML}{228B22}
\definecolor{bias}{HTML}{87CEFA}
\definecolor{noise}{HTML}{8B008B}
\definecolor{conv}{HTML}{FFA500}
\definecolor{pool}{HTML}{B22222}
\definecolor{up}{HTML}{B22222}
\definecolor{view}{HTML}{FFFFFF}
\definecolor{bn}{HTML}{FFD700}
\tikzset{fc/.style={black,draw=black,fill=fc,rectangle,minimum height=1cm}}
\tikzset{h/.style={black,draw=black,fill=h,rectangle,minimum height=1cm}}
\tikzset{bias/.style={black,draw=black,fill=bias,rectangle,minimum height=1cm}}
\tikzset{noise/.style={black,draw=black,fill=noise,rectangle,minimum height=1cm}}
\tikzset{conv/.style={black,draw=black,fill=conv,rectangle,minimum height=1cm}}
\tikzset{pool/.style={black,draw=black,fill=pool,rectangle,minimum height=1cm}}
\tikzset{up/.style={black,draw=black,fill=up,rectangle,minimum height=1cm}}
\tikzset{view/.style={black,draw=black,fill=view,rectangle,minimum height=1cm}}
\tikzset{bn/.style={black,draw=black,fill=bn,rectangle,minimum height=1cm}}
 
\usepackage{xspace}

\tikzstyle{dummy} = [coordinate]
\pgfkeys{/pgf/.cd,
  parallelepiped offset x/.initial=2mm,
  parallelepiped offset y/.initial=2mm
}
\pgfdeclareshape{parallelepiped}
{
  \inheritsavedanchors[from=rectangle] 
  \inheritanchorborder[from=rectangle]
  \inheritanchor[from=rectangle]{north}
  \inheritanchor[from=rectangle]{north west}
  \inheritanchor[from=rectangle]{north east}
  \inheritanchor[from=rectangle]{center}
  \inheritanchor[from=rectangle]{west}
  \inheritanchor[from=rectangle]{east}
  \inheritanchor[from=rectangle]{mid}
  \inheritanchor[from=rectangle]{mid west}
  \inheritanchor[from=rectangle]{mid east}
  \inheritanchor[from=rectangle]{base}
  \inheritanchor[from=rectangle]{base west}
  \inheritanchor[from=rectangle]{base east}
  \inheritanchor[from=rectangle]{south}
  \inheritanchor[from=rectangle]{south west}
  \inheritanchor[from=rectangle]{south east}
  \backgroundpath{
    \southwest \pgf@xa=\pgf@x \pgf@ya=\pgf@y
    \northeast \pgf@xb=\pgf@x \pgf@yb=\pgf@y
    \pgfmathsetlength\pgfutil@tempdima{\pgfkeysvalueof{/pgf/parallelepiped offset x}}
    \pgfmathsetlength\pgfutil@tempdimb{\pgfkeysvalueof{/pgf/parallelepiped offset y}}
    \def\ppd@offset{\pgfpoint{\pgfutil@tempdima}{\pgfutil@tempdimb}}
    \pgfpathmoveto{\pgfqpoint{\pgf@xa}{\pgf@ya}}
    \pgfpathlineto{\pgfqpoint{\pgf@xb}{\pgf@ya}}
    \pgfpathlineto{\pgfqpoint{\pgf@xb}{\pgf@yb}}
    \pgfpathlineto{\pgfqpoint{\pgf@xa}{\pgf@yb}}
    \pgfpathclose
    \pgfpathmoveto{\pgfqpoint{\pgf@xb}{\pgf@ya}}
    \pgfpathlineto{\pgfpointadd{\pgfpoint{\pgf@xb}{\pgf@ya}}{\ppd@offset}}
    \pgfpathlineto{\pgfpointadd{\pgfpoint{\pgf@xb}{\pgf@yb}}{\ppd@offset}}
    \pgfpathlineto{\pgfpointadd{\pgfpoint{\pgf@xa}{\pgf@yb}}{\ppd@offset}}
    \pgfpathlineto{\pgfqpoint{\pgf@xa}{\pgf@yb}}
    \pgfpathmoveto{\pgfqpoint{\pgf@xb}{\pgf@yb}}
    \pgfpathlineto{\pgfpointadd{\pgfpoint{\pgf@xb}{\pgf@yb}}{\ppd@offset}}
  }
}
\pgfdeclareshape{document}{
\inheritsavedanchors[from=rectangle] 
\inheritanchorborder[from=rectangle]
\inheritanchor[from=rectangle]{center}
\inheritanchor[from=rectangle]{north}
\inheritanchor[from=rectangle]{north east}
\inheritanchor[from=rectangle]{north west}
\inheritanchor[from=rectangle]{south}
\inheritanchor[from=rectangle]{south east}
\inheritanchor[from=rectangle]{south west}
\inheritanchor[from=rectangle]{west}
\inheritanchor[from=rectangle]{east}
\backgroundpath{%
\southwest \pgf@xa=\pgf@x \pgf@ya=\pgf@y
\northeast \pgf@xb=\pgf@x \pgf@yb=\pgf@y
\pgf@xc=\pgf@xb \advance\pgf@xc by-5pt 
\pgf@yc=\pgf@ya \advance\pgf@yc by5pt
\pgfpathmoveto{\pgfpoint{\pgf@xa}{\pgf@ya}}
\pgfpathlineto{\pgfpoint{\pgf@xa}{\pgf@yb}}
\pgfpathlineto{\pgfpoint{\pgf@xb}{\pgf@yb}}
\pgfpathlineto{\pgfpoint{\pgf@xb}{\pgf@yc}}
\pgfpathlineto{\pgfpoint{\pgf@xc}{\pgf@ya}}
\pgfpathclose
\pgfpathmoveto{\pgfpoint{\pgf@xc}{\pgf@ya}}
\pgfpathlineto{\pgfpoint{\pgf@xc}{\pgf@yc}}
\pgfpathlineto{\pgfpoint{\pgf@xb}{\pgf@yc}}
\pgfpathclose
}
}
\tikzstyle{block} = [draw, fill=white, rectangle, minimum height=3em, minimum width=6em]
    
\usetikzlibrary{backgrounds}
\usepackage{pifont}
\tikzstyle{startstop} = [rectangle, rounded corners, minimum width=0.8cm, minimum height=0.8cm,text centered, draw=black, fill=lime!30]
\usetikzlibrary{chains}
\usepackage{booktabs}
\usepackage{multirow}
\usepackage[margin=0.75in]{geometry}
\usepackage{graphicx}
\usepackage{amsmath}
\usepackage{xcolor}
\usepackage{tikz}
\usepackage{caption}
\usepackage{subcaption}
\usepackage{microtype}
\usetikzlibrary{arrows.meta,backgrounds,calc,fit,positioning}

\definecolor{roomfill}{HTML}{F7F8FA}
\definecolor{wallcolor}{HTML}{263238}
\definecolor{directcolor}{HTML}{E64A19}
\definecolor{multipathcolor}{HTML}{2878B5}
\definecolor{arraycolor}{HTML}{2E7D32}
\definecolor{personcolor}{HTML}{6A1B9A}
\definecolor{deskcolor}{HTML}{B98555}
\definecolor{tablecolor}{HTML}{8D6E63}
\definecolor{sourceblue}{HTML}{2F6BFF}
\definecolor{noisegray}{HTML}{667085}
\definecolor{physicsorange}{HTML}{E64A19}
\definecolor{roomorange}{HTML}{FFF0D8}
\definecolor{networkpurple}{HTML}{EEE8FF}
\definecolor{headgreen}{HTML}{E1F5E9}
\definecolor{kdred}{HTML}{D92D20}
\definecolor{ink}{HTML}{172B4D}
\definecolor{softline}{HTML}{98A2B3}

\newcommand{\lossSELD}{\mathcal{L}_{\mathrm{SELD}}} 
\newcommand{\lossKD}{\mathcal{L}_{\mathrm{PGKD}}} 

\newcommand\MakeUppercaseGreek[1]{
  \begingroup
    \let\psi\Psi
    \let\omega\Omega
    \let\gamma\Gamma
    \MakeUppercase{#1}
  \endgroup}

\newcommand{\vectorsym}[1]{\bm{#1}}

\newcommand{\brackets}[1]{\left(#1\right)}

\newcommand{\matsym}[1]{\mathbf{#1}}
\newcommand{\squareb}[1]{\left[{#1}\right]}

\newcommand{\clip}[3]{\mathrm{clip}\left(#1,#2,#3\right)}
\newcommand{\xmark}{\ding{55}}%
\newcommand{\cmark}{\ding{51}}%

\newcommand{\p}[0]{\vectorsym{\theta}}

\newcommand{\z}[0]{\vectorsym{z}}
\newcommand{\x}[0]{\vectorsym{x}}

\newcommand{\method}{PG-SELD}

\title{PG-SELD: PHYSICS-GUIDED SOUND EVENT LOCALIZATION AND DETECTION}

\name{
Elad Cohen \orcidlink{0009-0003-7724-2323},
Elad Dror Cohen \orcidlink{0009-0001-1868-9243},
Arnon Netzer \orcidlink{0009-0000-5339-9439},
Hai Victor Habi \orcidlink{0000-0001-5612-6348}
}

\address{Arm Holdings, Israel}

\begin{document}

\maketitle

\begin{abstract}
Sound event localization and detection (SELD) aims to jointly recognize sound events and estimate their directions of arrival from multichannel audio. Although recent deep learning approaches have achieved strong performance, their ability to generalize across acoustic environments remains limited, as room reverberation introduces environment-specific characteristics into the learned representations. In this work, we address this challenge by leveraging a physical free-field model as a room-independent reference. Specifically, we propose \method{}, a training framework that combines free-field with physics-guided knowledge distillation. Our approach aligns intermediate representations extracted from reverberant signals with those produced by a free-field teacher for matched acoustic scenes. This guidance encourages the model to preserve event- and localization-relevant information while reducing sensitivity to room-specific characteristics. Experimental results on the STARSS23 benchmark show that \method{} consistently improves the generalization performance of multiple baseline SELD architectures.
\end{abstract}

\begin{keywords}
SELD, DOA, RIR, Knowledge Distillation
\end{keywords}

\section{Introduction}

Sound event localization and detection (SELD) \cite{berghi2024fusion, adavanne2019seldnet,hu2025pseldnets} jointly estimates the temporal activity and spatial location of sound events from multichannel audio by predicting their classes and directions of arrival (DOAs). This unified framework enables spatially aware acoustic scene analysis for applications such as surveillance, teleconferencing, assistive hearing systems, and mobile robotics operating across changing acoustic environments. Recent deep-learning architectures, including convolutional-recurrent, Conformer, and Transformer-based models, have substantially improved SELD performance by learning joint semantic and spatial representations \cite{berghi2024fusion,adavanne2019seldnet,hu2025pseldnets,perotin2019crnn}. Despite these advances, generalization across acoustic environments remains challenging. Reverberation, noise, microphone characteristics, and room geometry distort localization cues, leading models to learn room-dependent features that degrade in unseen environments. Moreover, collecting real multichannel recordings with accurate spatial annotations is costly \cite{shimada2023starss23}. SELD systems therefore rely heavily on synthetic datasets generated using room acoustic simulators or measured SRIRs, which provide controlled acoustic conditions and accurate annotations \cite{scheibler2018pyroomacoustics}. However, such data may encourage models to learn rendering- or room-specific characteristics rather than robust acoustic representations. Existing approaches mitigate this mismatch through data augmentation, including spatial transformations, channel manipulations, frequency perturbations, and signal mixing \cite{wang2023fourstage}. Multi-room simulation further increases diversity by varying room geometry, reverberation, and source configurations. Other approaches combine synthetic and real recordings or leverage large-scale pretraining \cite{hu2025pseldnets,shimada2023starss23,fonseca2022fsd50k}. While these strategies improve robustness, they do not explicitly enforce room-robust representations, leaving models sensitive to room-specific acoustic signatures. Recent works have addressed environmental robustness in SELD through
echo-aware feature adaptation \cite{yasuda2022echo}, physically informed spatial regularization \cite{liu2025physically}, and intermediate-feature distillation for low-resource audio-visual SELD \cite{wang2026hda}.

In this paper, we propose \method{}, a physics-guided SELD approach that encourages room-robust representations. \method{} leverages free-field recordings as a room-independent reference to guide representation learning under reverberant conditions. By aligning reverberant representations with their free-field counterparts through physics-guided feature regularization, the model is encouraged to retain localization- and event-relevant acoustic cues while reducing sensitivity to room-specific characteristics. \method{} is model-agnostic, introduces no additional inference cost, and can be readily integrated into existing SELD architectures. We evaluate \method{} on CNN-Conformer \cite{berghi2024fusion}, SELDNet \cite{adavanne2019seldnet}, and HTS-AT \cite{chen2022htsat}, demonstrating consistent improvements in generalization performance across all three architectures.

Our main contributions are summarized as follows:
\begin{itemize}
\item We propose \method{}, a physics-guided training framework that exploits free-field renderings to learn room-robust representations for sound event localization and detection.
\item We introduce a physics-guided knowledge distillation (PGKD) regularization that encourages room-robust latent representations by aligning features extracted from matched reverberant and free-field recordings.
\item We demonstrate that the proposed training strategy consistently improves the generalization performance of multiple baseline SELD architectures on real acoustic environments.
\end{itemize}


\section{Signal Model and Notation}

Here, we begin by defining the response of a microphone array to an incident acoustic signal. Specifically, we consider an array of \(M\) microphones receiving signals from \(Q\) sound sources. The corresponding multichannel signal, represented in the short-time Fourier transform (STFT) domain, can then be expressed as \cite{benesty2008microphone}: 
\begin{align}\label{eq:signal_gen}
\vectorsym{X}(t,f)&=\sum_{q=1}^{Q}H_q(f)S_q(t,f) + \vectorsym{V}(t,f)\nonumber\\
&=\matsym{H}(f)\vectorsym{S}(t,f)+\vectorsym{V}(t,f),
\end{align}
where \(t\) and \(f\) denote the time-frame and frequency, 
respectively. Here, \(S_q(t,f)\) represents the STFT coefficient of the \(q\)-th
source in the time-frequency bin \((t,f)\), \(H_q(f)\) denotes the acoustic transfer function between the \(q\)-th source and the microphone array, and \(\vectorsym{V}\) denotes additive noise. In free-field environments, the far-field transfer function between the $q$-th source and the $m$-th microphone is modeled as
\begin{align}\label{eq:freefield}
\squareb{H_q(f)}_m=A\exp\!\left[-j\,2\pi f\tau_m(\phi_q,\theta_q)\right],
\end{align}
where $\tau_m(\phi_q,\theta_q)$ is the relative propagation delay at the $m$-th microphone with respect to a reference microphone, $\phi_q$ and $\theta_q$ are azimuth and elevation of the \(q\)-th source, respectively, and $A$ is a constant amplitude factor. In reverberant environments, $\squareb{H_q(f)}_m$ contains both direct-path propagation and room-dependent reflections, which can distort the spatial cues required for accurate sound source localization. In contrast, free-field acoustics preserve predominantly direct-path information, providing cleaner spatial representations that are less affected by room-specific characteristics. 
\section{Method}

\begin{figure}[!t]
    \centering
\begingroup
    \setlength{\fboxsep}{0pt}%
    \resizebox{\columnwidth}{!}{%
\colorbox{white}{%
\begin{tikzpicture}[
  font=\sffamily,
  >={Latex[length=2.8mm,width=1.8mm]},
  flow/.style={->, draw=ink, line width=0.9pt, rounded corners=2pt},
  signal/.style={flow, draw=sourceblue},
  process/.style={
    draw=softline, rounded corners=3pt, minimum height=13mm,
    align=center, font=\sffamily\small, line width=0.8pt
  },
  branchmodel/.style={process, text width=30mm, inner xsep=1.5mm},
  database/.style={
    cylinder, shape border rotate=90, aspect=0.25,
    minimum width=30mm, minimum height=13mm,
    align=center, font=\sffamily\small, line width=0.8pt,
    cylinder uses custom fill
  },
  preprocess/.style={
    process, fill=white, minimum width=20mm, minimum height=10mm,
    font=\sffamily\scriptsize
  },
  encoder/.style={process, fill=networkpurple, minimum width=19mm},
  backbone/.style={process, fill=networkpurple!72, minimum width=22mm},
  head/.style={
    process, fill=headgreen, minimum width=19mm, minimum height=13mm,
    font=\sffamily\scriptsize
  },
  group/.style={draw=softline, rounded corners=5pt, line width=0.8pt},
  kd/.style={
    ->, draw=kdred, dashed, dash pattern=on 3pt off 2pt,
    line width=1.15pt
  },
  label/.style={font=\sffamily\scriptsize, text=ink, align=center},
  fireicon/.pic={
    \path[draw=kdred!70!black, fill=kdred!88, line width=0.25pt]
      (0,-0.12) .. controls (-0.12,-0.02) and (-0.05,0.10) .. (0.02,0.17)
      .. controls (0.02,0.06) and (0.15,0.03) .. (0.11,0.18)
      .. controls (0.26,0.05) and (0.20,-0.10) .. (0.06,-0.17)
      .. controls (0.03,-0.18) and (-0.03,-0.18) .. (0,-0.12);
    \path[fill=physicsorange!70]
      (0.02,-0.12) .. controls (-0.04,-0.04) and (0.04,0.04) .. (0.06,0.10)
      .. controls (0.13,0.02) and (0.14,-0.09) .. (0.02,-0.12);
  },
  iceicon/.pic={
    \foreach \a in {0,60,...,300}{
      \draw[sourceblue!75!white, line width=0.35pt, rotate=\a] (0,0) -- (0,0.18);
      \draw[sourceblue!75!white, line width=0.25pt, rotate=\a] (0,0.11) -- (0.035,0.145);
      \draw[sourceblue!75!white, line width=0.25pt, rotate=\a] (0,0.11) -- (-0.035,0.145);
    }
    \fill[sourceblue!25!white] (0,0) circle (0.025);
  }
]

\node[branchmodel, draw=physicsorange!55!black, fill=physicsorange!10]
  (freespace) at (4.80,1.65)
  {Free-Field\\Acoustic Model};
\node[preprocess] (prepT) at (9.20,1.65) {\large Preprocessing};
\node[backbone] (backT) at (12.80,1.65) {\large Backbone};
\node[head] (headT) at (16.80,1.65) {\large Head};

\draw[signal] (freespace) -- node[midway, above=1pt, label, fill=white, inner sep=1pt]
  {\large $\x_{ff}$} (prepT);
\draw[flow] (prepT) -- (backT);
\draw[flow] (backT) -- node[midway, above=1pt, label, fill=white, inner sep=1pt]
  {\large $\z_{ff}$} (headT);
\coordinate (outT) at ($(headT.east)+(10mm,0)$);
\draw[flow] (headT.east) -- (outT);

\path ($(backT.north east)+(-3.0mm,-2.2mm)$) pic[scale=0.55] {iceicon};
\path ($(headT.north east)+(-3.0mm,-2.2mm)$) pic[scale=0.50] {iceicon};

\node[database, draw=roomorange!55!black,
  cylinder body fill=roomorange, cylinder end fill=roomorange!70]
  (synthesis) at (4.80,-1.65)
  {\Large $\mathcal{D}_{RIR}$};
\node[preprocess] (prepS) at (9.20,-1.65) {\large Preprocessing};
\node[backbone] (backS) at (12.80,-1.65) {\large Backbone};
\node[head] (headS) at (16.80,-1.65) {\large Head};

\draw[signal] (synthesis.north) --
  node[midway, right=2pt, label, fill=white, inner sep=1pt]
  {Sources and\\DOA labels} (freespace.south);
\draw[signal] (synthesis) -- node[midway, above=1pt, label, fill=white, inner sep=1pt]
  {\large $\x_{rev}$} (prepS);
\draw[flow] (prepS) -- (backS);
\draw[flow] (backS) -- node[midway, below=1pt, label, fill=white, inner sep=1pt]
  {\large $\z_{rev}$} (headS);
\coordinate (outS) at ($(headS.east)+(10mm,0)$);
\draw[flow] (headS.east) -- (outS);
\node[label, anchor=west, text=kdred, inner sep=1pt]
  at ([xshift=1.5mm]outS) {\large $\lossSELD$};

\path ($(backS.north east)+(-3.0mm,-2.2mm)$) pic[scale=0.52] {fireicon};
\path ($(headS.north east)+(-3.0mm,-2.2mm)$) pic[scale=0.48] {fireicon};

\coordinate (kdT) at ($(backT.east)!0.50!(headT.west)$);
\coordinate (kdS) at ($(backS.east)!0.50!(headS.west)$);
\fill[kdred] (kdT) circle (1.1pt);
\fill[kdred] (kdS) circle (1.1pt);
\draw[kd] (kdT) -- node[midway, label, fill=white, text=kdred, inner sep=1pt]
  {\large $\lossKD$} (kdS);

\begin{scope}[on background layer]
  \node[group, draw=physicsorange!45!black, fill=physicsorange!8,
    fit=(freespace)(prepT)(backT)(headT),
    inner xsep=2.5mm, inner ysep=4mm] (teacherbox) {};
  \node[group, fill=roomorange!20,
    fit=(synthesis)(prepS)(backS)(headS),
    inner xsep=2.5mm, inner ysep=4mm] (studentbox) {};
\end{scope}
\node[anchor=south, font=\sffamily\bfseries\small, text=ink]
  at ([yshift=1.5mm]teacherbox.north)
  {\large Free-Field Teacher};
\node[anchor=south, font=\sffamily\bfseries\small, text=ink]
  at ([yshift=1.5mm]studentbox.north)
  {\large Reverberant Student};

\end{tikzpicture}
}%
    }
\endgroup
    \caption{PGKD: The free-field teacher and reverberant student with
    backbone-level knowledge distillation.}
    \label{fig:teacher-student}
\end{figure}
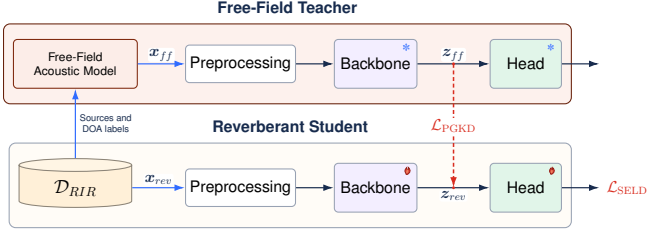

Our objective is to learn a neural network that remains robust under changing environmental conditions. To achieve this, we propose \method{}, a physics-guided SELD approach that leverages the free-field model to learn room-robust representations for SELD. Our approach is composed of two main components: (i) a Physics-Guided Knowledge Distillation (PGKD) method described in Section~\ref{PGKD} and illustrated in Fig. \ref{fig:teacher-student}; and (ii) a training strategy that exploits the free-field model as a training objective, detailed in Section~\ref{training}.

\subsection{Physics-Guided Knowledge Distillation (PGKD)}\label{PGKD}
In order to provide room-independent features, we propose Physics-Guided Knowledge Distillation (PGKD) by utilizing the free-field array response from \eqref{eq:freefield}. This concept builds on the physics-guided approach \cite{faroughi2022physics}, in which physical knowledge is infused into the model through the generation of training data. In our work, we extend this idea by using it to regulate the model’s intermediate representations via knowledge distillation (KD) \cite{romero2015fitnets} and by embedding this information directly into the model.
\begin{figure}[t]
\centering
\includesvg[width=0.8\linewidth]{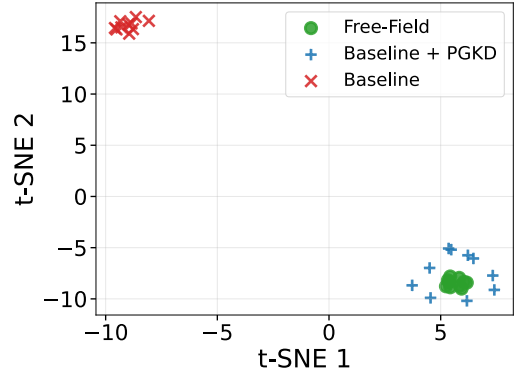}
\caption{t-SNE visualization of SELDNet \cite{adavanne2019seldnet} features. Samples are taken from the pairs ($\vectorsym{x}_{rev}$, $\vectorsym{x}_{ff}$).}
\label{fig:tsne}
\end{figure}
Specifically, we are given an acoustic scene dataset $\mathcal{D}_{RIR}=\{\matsym{H}_n,\vectorsym{s}_n,\mathcal{Y}_n\}_{n=1}^N$, where each sample contains room impulse responses (RIR) $\matsym{H}_n$,  $\vectorsym{s}$ source signals, and labels $\mathcal{Y}$ which contain class labels, temporal activity, and DOAs. Using the dataset $\mathcal{D}_{RIR}$ with the free-field signal modeling in \eqref{eq:freefield}, we construct a pair of inputs, $\vectorsym{x}_{rev}$ and $\vectorsym{x}_{ff}$, which represent the array responses of the same sources in a reverberant environment and in free field, respectively. Since each pair ($\vectorsym{x}_{rev}$, $\vectorsym{x}_{ff}$) shares the same source signals, event activity, and spatial configuration, but differs in acoustic propagation conditions, the free-field representation provides a room-independent target for the corresponding reverberant input. 

After constructing these sample pairs, we define the teacher and student models. Let $\vectorsym{f}_{l}$ denote the $l$-th feature representation of the network to be trained for robust SELD (student), parameterized by $\Omega$. Similarly, let $\vectorsym{g}_{l}$ denote the corresponding representation extracted from a teacher model, parameterized by $\Omega_g$, which shares the same architecture as the student and is pretrained on $\mathcal{D}_{G}$, a large-scale free-field dataset. $\mathcal{D}_{G}$ is synthesized from \eqref{eq:freefield} using sources $\vectorsym{s}$ across all directions. The teacher provides reference representations that guide the learning of the student. Specifically, we encourage the student to align these representations from reverberant inputs by introducing a feature-level distillation objective based on cosine similarity:
\begin{align}    &\vectorsym{z}_{rev}\brackets{\Omega}=\vectorsym{f}_{l}\brackets{\vectorsym{x}_{rev};\Omega}, \quad\text{and}\quad
\vectorsym{z}_{ff}=\vectorsym{g}_{l}\brackets{\vectorsym{x}_{ff};\Omega_g}\nonumber\\
&\mathcal{L}_{\mathrm{PGKD}}\brackets{\Omega,\matsym{W}}=1-\frac{(\matsym{W}\z_{rev})^\top \z_{ff}}{\|\matsym{W}\z_{rev}\|_2\,\|\z_{ff}\|_2},
\end{align}
where $\matsym{W}$ is a projection matrix. $\matsym{W}$ corrects feature misalignment between teacher and student \cite{miles2024projector}, whose latent coordinate systems may differ due to being optimized independently under different acoustic conditions, and is discarded at inference. Feature alignment is performed on the backbone output, immediately before the prediction head. At this level, the representations encode higher-level semantic and spatial information and are less dominated by low-level acoustic variations. The teacher parameters $\Omega_g$ remain frozen during this phase, and $\vectorsym{z}_{ff}$ is kept detached. 
Minimizing $\mathcal{L}_{\mathrm{PGKD}}$ encourages the student to produce representations that are consistent with those obtained under free-field conditions, thus reducing the sensitivity to room-dependent reverberation while retaining information relevant to SELD. In Fig.~\ref{fig:tsne}, we present a t-SNE \cite{van2008visualizing} visualization of the feature representations under the PGKD loss, showing that the baseline model learns distinctly different features, whereas PGKD encourages the features to be closer to those of the free-field conditions.

\subsection{Training Strategy}\label{training}

To further enhance the model’s capabilities, we also employ physics-guided training using the free-field model. We adopt the multi-activity Cartesian coordinate direction-of-arrival (multi-ACCDOA) \cite{shimada2022multiaccdoa} with 3 tracks, utilizing the auxiliary duplicating permutation invariant training (PIT) \cite{yu2017pit} framework with the mean squared loss as the task loss function. Then, the student is trained on $\mathcal{D}_{{RIR}}$ using PGKD. Each training sample contains a paired reverberant input $\vectorsym{x}_{{rev}}$ and free-field input $\vectorsym{x}_{{ff}}$; the frozen teacher processes $\vectorsym{x}_{{ff}}$, while the student input is selected probabilistically following a curriculum learning strategy. This strategy gradually shifts the student’s inputs from exclusively $\vectorsym{x}_{{ff}}$ to exclusively $\vectorsym{x}_{{rev}}$, enabling the model to first learn under controlled acoustic conditions and then progressively adapt to realistic reverberant recordings. Note that the student never directly observes $\mathcal{D}_{G}$; knowledge from the teacher is transferred solely through the feature-level regularization term. The general objective for training the student is given by
\begin{align}
\mathcal{L}\brackets{\Omega,\matsym{W}}=\mathcal{L}_{\mathrm{SELD}}\brackets{\Omega}+\lambda\mathcal{L}_{\mathrm{PGKD}}\brackets{\Omega,\matsym{W}},
\end{align}
where \(\mathcal{L}_{\mathrm{SELD}}\) is the standard SELD loss and \(\lambda\) is the distillation weight, and both losses are computed frame-wise and averaged over time.

\section{Experimental Results}
In this section, we evaluate \method{} on the following SELD architectures: CNN-Conformer \cite{berghi2024fusion}, SELDNet \cite{adavanne2019seldnet} and HTS-AT \cite{chen2022htsat}. We first describe the experimental setup and implementation details, followed by a comparison with conventional training strategies. Finally, we conduct ablation studies to analyze the contribution of each component and the robustness of PGKD to varying room diversity.

\subsection{Implementation Details}
\noindent\textbf{Datasets:} We construct a free-field-style dataset $\mathcal{D}_{G}$, which is used to pretrain the teacher. To synthesize this dataset, we rely on FSD50K \cite{fonseca2022fsd50k}, from which we extract events that match the DCASE target classes \cite{shimada2023starss23}. Then, we synthesize free-field recordings with azimuth and elevation uniformly sampled from $\phi\in[-180^\circ,179^\circ]$ and $\theta\in[-80^\circ,80^\circ]$, respectively. Next, to build the paired dataset of PGKD for student training, we employ the spatial room impulse response (SRIR) database TAU-SRIR-DB \cite{politis2022tausrirdb} in combination with FSD50K \cite{fonseca2022fsd50k}, and generate paired free-field and reverberant recordings for each room, using six rooms for training and three for validation. All scenes were generated similarly to \cite{hu2025pseldnets} and are one minute long with a maximum of $ Q=3$ sources. Finally, we use real recordings from the STARSS23 \cite{shimada2023starss23} dataset for fine-tuning. Table~\ref{table:setup} summarizes the datasets used for the teacher and student.

\begin{table}[t]
\centering
\resizebox{0.49\textwidth}{!}{
\begin{tabular}{llll}
\toprule
\rowcolor{gray!20}
\textbf{Model} & \textbf{Teacher} 
               & \multicolumn{2}{c}{\textbf{Student}} \\
\rowcolor{gray!20}
               & 
               & \textbf{\method{}} & \textbf{Fine-Tune} \\
\midrule
Dataset Type 
    & Synthetic 
    & Synthetic 
    & Real \\

Acoustic condition 
    & Free-Field 
    & Free-Field, SRIR 
    & Real-room \\

Audio Source 
    & \cite{fonseca2022fsd50k} 
    & \cite{fonseca2022fsd50k,politis2022tausrirdb} 
    & \cite{shimada2023starss23} \\

Training Set [hours] 
    & 120 
    & 15 (paired) 
    & 3.8 \\

Validation Set [hours] 
    & 30 
    & 5 
    & 3.2 \\
\bottomrule
\end{tabular}}
\caption{Datasets.}
\label{table:setup}
\end{table}

\noindent\textbf{Preprocessing.}\indent Following signal generation in the microphone domain according to \eqref{eq:signal_gen} and \eqref{eq:freefield}, both the free-field and reverberant recordings are converted to four-channel First-Order Ambisonics (FOA) at 24~kHz before feature extraction. Then, we extract features by concatenating 4-channel log-mel spectrogram features with 3-channel normalized acoustic intensity vectors \cite{perotin2019crnn}. During training, recordings are segmented into fixed-length clips with 50\% overlap, while during inference, no overlap is used.

\noindent\textbf{Training.}\indent In all experiments, we adopt the following setup and hyperparameters unless stated otherwise. We train the models with a batch size of 64 for 100 epochs using the AdamW optimizer, starting from the base learning rate of $10^{-4}$ for CNN-Conformer, $3\cdot10^{-4}$ for HTS-AT, and $10^{-3}$ for SELDNet. Then, the learning rate was reduced by a factor of 10 during the final 10 epochs. During training, we apply the following augmentations: Audio Channel Swapping (ACS) \cite{wang2023fourstage}, SpecAugment \cite{park2019specaugment}, and frequency shifting. During PGKD, we set $\lambda = 0.5$, and all stochastic augmentations are synchronized within each matched pair. Finally, we fine-tune the models for an additional 30 epochs on the STARSS23 train split, using $0.1$ times the base learning rate.

\begin{table}[]
\centering
\small
\resizebox{0.49\textwidth}{!}{
\begin{tabular}{lcccccc}
\toprule
\rowcolor{gray!20}
Model           &  \textbf{\method{}}  & $ER_{20^\circ}\downarrow$  & $F_{20^\circ}\uparrow$ & $LE_{\mathrm{CD}}\downarrow$ & $LR_{\mathrm{CD}}\uparrow$ & $\mathcal{E}_{\mathrm{SELD}}\downarrow$ \\ \midrule
CNN-Conformer \cite{berghi2024fusion}     &  \xmark   &  0.502  & 55.6\%  &  13.7\degree  &  61.9\%  &  0.350   \\
&  \cmark    &  0.449  & 61.9\%  &  12.2\degree &  67.2\%  &  \textbf{0.306}    \\ \midrule
SELDNet \cite{adavanne2019seldnet} &   \xmark   &  0.543  &46.1\%  &  29.2\degree  &  53.2\%  &  0.428    \\
& \cmark    &  0.529  & 47.2\%   &  17.1\degree  &  54.7\%   &  \textbf{0.400}    \\ \midrule
HTS-AT \cite{chen2022htsat} &   \xmark   &  0.538  & 50.9\%  &  14.5\degree  &  56.8\%  &  0.385   \\
&  \cmark    &  0.519  & 54.2\%  &  13.6\degree  &  59.9\%  &  \textbf{0.363}    \\ \bottomrule
\end{tabular}}
\caption{SELD performance comparison on the validation split of $\mathcal{D}_{RIR}$.}
\label{table:main_synth}
\end{table}

\begin{table}[]
\centering
\small
\resizebox{0.49\textwidth}{!}{
\begin{tabular}{lcccccc}
\toprule
\rowcolor{gray!20}
Model           &  \textbf{\method{}}  & $ER_{20^\circ}\downarrow$  & $F_{20^\circ}\uparrow$ & $LE_{\mathrm{CD}}\downarrow$ & $LR_{\mathrm{CD}}\uparrow$ & $\mathcal{E}_{\mathrm{SELD}}\downarrow$ \\ \midrule
CNN-Conformer \cite{berghi2024fusion}     &  \xmark   &  0.544  & 45.4\%  &  17.5\degree  &  65.0\%  &  0.384   \\
&  \cmark    &  0.514  & 50.5\%  &  15.4\degree &  65.1\%  &  \textbf{0.361}    \\ \midrule
SELDNet \cite{adavanne2019seldnet} &   \xmark   &  0.567  & 37.3\%  &  31.4\degree  &  56.5\%  &  0.451    \\
& \cmark    &  0.562  & 38.4\%   &  20.0\degree  &  64.8\%   &  \textbf{0.410}    \\ \midrule
HTS-AT \cite{chen2022htsat} &   \xmark   &  0.562  & 47.6\%   &  15.8\degree  &  62.1\%   &  0.388    \\
&  \cmark    &  0.548  & 48.2\%  &  17.0\degree  &  67.0\%  &  \textbf{0.373}    \\ \bottomrule
\end{tabular}}
\caption{SELD performance comparison on the test split of STARSS23 \cite{shimada2023starss23}.}
\label{table:main}
\end{table}

\noindent\textbf{Evaluation.}\indent We evaluated the performance on the validation set of $\mathcal{D}_{RIR}$ and on the test split of STARSS23 \cite{shimada2023starss23}. SELD performance is evaluated using the standard joint localization and detection metrics with macro-averaging \cite{mesaros2019joint}: location-dependent detection metrics F-score ($F_{20^\circ}$) and error rate ($ER_{20^\circ}$), together with class-dependent localization metrics localization recall ($LR_{\mathrm{CD}}$) and localization error ($LE_{\mathrm{CD}}$). For model comparison and hyperparameter selection, we use the aggregated SELD metric
$\mathcal{E}_{\mathrm{SELD}} = \frac{1}{4}\left(ER_{20^\circ} + (1-F_{20^\circ}) + \frac{LE_{\mathrm{CD}}}{180^\circ} + (1-LR_{\mathrm{CD}})\right)$,
which equally weights the four SELD evaluation metrics.

\begin{table}[t]
\centering
\begin{minipage}[t]{0.26\textwidth}
\centering
\begin{tabular}{ccc}
\toprule
\rowcolor{gray!20}
PGKD & Curriculum &
$\mathcal{E}_{\mathrm{SELD}}\downarrow$ \\
\midrule
\xmark & \xmark & 0.350 \\
\cmark & \xmark & 0.318 \\
\cmark & \cmark & 0.306 \\
\bottomrule
\end{tabular}
\caption*{(a)}
\end{minipage}
\hfill
\begin{minipage}[t]{0.20\textwidth}
\centering
\begin{tabular}{lc}
\toprule
\rowcolor{gray!20}
$\lambda$ & $\mathcal{E}_{\mathrm{SELD}}\downarrow$ \\
\midrule
0.1 & 0.312 \\
0.5 & 0.306 \\
1.0 & 0.314 \\
\bottomrule
\end{tabular}
\caption*{(b)}
\end{minipage}
\caption{(a) \method{} Components. (b) PGKD $\lambda$ ablation.}
\label{table:ablation}
\end{table}

\begin{figure}[t]
\centering
\includesvg[width=0.8\linewidth]{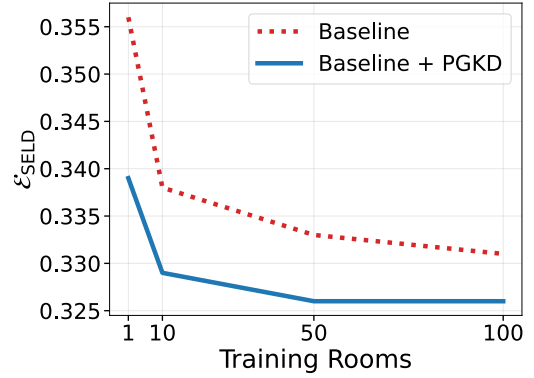}
\caption{Number of training rooms vs. $\mathcal{E}_{\mathrm{SELD}}$}
\label{fig:rooms}
\end{figure}

\subsection{Results}
We compare \method{} with conventional synthetic SRIR pre-training followed by STARSS23 fine-tuning  \cite{berghi2024fusion,hu2025pseldnets}, using identical $\mathcal{D}_{RIR}$, optimization, and augmentations, differing only in PGKD. Table~\ref{table:main_synth} compares SELD performance with and without \method{} across different models on the validation set of $\mathcal{D}_{RIR}$. Our approach improves all evaluation metrics, demonstrating its effectiveness across diverse model architectures. The gains are primarily driven by improved localization, consistent with the spatial nature of free-field guidance. Table~\ref{table:main} shows consistent reductions in $\mathcal{E}_{\mathrm{SELD}}$ on STARSS23 for all baseline models, indicating that the improvements extend beyond synthetic data to real-world recordings. The observed SELD improvements across architectures indicate that \method{} is not tied to a specific backbone design.

\subsection{Ablation}

\noindent\textbf{\method{} Components.}\indent Table~\ref{table:ablation} presents an ablation study of the main components of \method{} using CNN-Conformer \cite{berghi2024fusion} on the validation set of $\mathcal{D}_{RIR}$. As shown in Table~\ref{table:ablation}(a), PGKD substantially improves SELD performance over the baseline, while incorporating the curriculum learning strategy provides an additional gain, demonstrating the benefit of progressively transitioning from free-field to reverberant inputs. Table~\ref{table:ablation}(b) further examines the PGKD distillation weight $\lambda$, with the best performance obtained at $\lambda=0.5$, which is therefore used in all our experiments.

\noindent\textbf{PGKD and Number of Rooms.}\indent To study the effect of room diversity, we generate a separate reverberant dataset across multiple simulated training rooms using Pyroomacoustics \cite{scheibler2018pyroomacoustics} with 300 rooms for validation. We evaluate CNN-Conformer \cite{berghi2024fusion} with and without PGKD while keeping the amount of 15 hours of data fixed and only varying the number of training rooms. Fig.~\ref{fig:rooms} shows that PGKD provides larger gains in the lower-room-diversity regime, while its additional benefit decreases as the training set becomes more diverse. This trend further suggests that PGKD improves robustness to room-specific acoustics by guiding the model toward features that are less dependent on individual room characteristics, whereas the baseline is more prone to exploiting room-specific information when room diversity is limited.

\noindent\textbf{t-SNE Feature Visualization.}\indent In Fig.~\ref{fig:tsne}, we visualize the pre-head feature representations of SELDNet \cite{adavanne2019seldnet} using t-SNE \cite{van2008visualizing}, where we apply only PGKD regularization. The features learned by PGKD exhibit greater similarity to those of the free-field model compared with the baseline, suggesting that PGKD guides the SELD model toward representations that better preserve free-field acoustic characteristics and reduce sensitivity to room acoustics.

\section{Conclusions}
In this paper, we proposed \method{}, a physics-guided SELD that leverages paired reverberant and free-field renderings to encourage room-robust representations. By aligning the latent features of matched recordings, the proposed method encourages the network to preserve event- and localization-related information while suppressing room-dependent acoustic characteristics. Experimental results on the STARSS23 benchmark demonstrate consistent improvements across multiple baseline models, highlighting the effectiveness of physics-guided representation learning for improving generalization to unseen acoustic environments.

\newpage

\bibliographystyle{IEEEbib}
\bibliography{strings,refs}

\end{document}